# A Persian ASR-based SER: Modification of Sharif Emotional Speech Database and Investigation of Persian Text Corpora

Ali Yazdani
Faculty of Computer Science and Engineering
Shahid Beheshti University
Tehran, Iran
ali.yazdani@mail.sbu.ac.ir

Yasser Shekofteh
Faculty of Computer Science and Engineering
Shahid Beheshti University
Tehran, Iran
y_shekofteh@sbu.ac.ir

*Abstract*—Speech Emotion Recognition (SER) is one of the essential perceptual methods of humans in understanding the situation and how to interact with others, therefore, in recent years, it has been tried to add the ability to recognize emotions to human-machine communication systems. Since the SER process relies on labeled data, databases are essential for it. Incomplete, low-quality or defective data may lead to inaccurate predictions. In this paper, we fixed the inconsistencies in Sharif Emotional Speech Database (ShEMO), as a Persian database, by using an Automatic Speech Recognition (ASR) system and investigating the effect of Farsi language models obtained from accessible Persian text corpora. We also introduced a Persian/Farsi ASR-based SER system that uses linguistic features of the ASR outputs and Deep Learning-based models.

## I. INTRODUCTION

The emotional state of humans is an important factor in their interactions and affects most communication channels such as facial expressions, voice characteristics, and the linguistic content of verbal communication. Speech is one of the main ways for expressing emotions, and therefore for a natural Human-Computer Interaction (HCI) system, recognizing, interpreting, and responding to emotions expressed in the speech is important[1]–[4]. The emotions, e.g. fear or anger, affect both the acoustic characteristics and the linguistic content of the speech[2], [5].

The speech emotion recognition (SER) systems aim to facilitate the natural interaction of humans with machines through direct voice interaction instead of using traditional devices as the input to understand the verbal content and ease the reaction of human listeners. Many problems in HCI systems need to be properly addressed, especially when these systems move from the lab environment to real-time applications[4], [6].

Since linguistic information can also be derived from speech, we can combine the acoustic features of the speech with linguistic information. Recent studies confirm that multi-modal systems perform better than unimodal emotion recognition systems. Multimodal emotion recognition shows significant performance improvement by fusing acoustic and linguistic information[5], [7]–[10].

It should be noted that as far as we know, a reliable system with proper performance for recognizing emotion in the Persian language has not been reported so far. Our main goal is to use the output textual information of a Persian Automatic Speech Recognition (ASR) system with a suitable language model (LM) in SER. Also, using the Persian ASR system, the ShEMO dataset has been modified. The ShEMO dataset is a Persian SER dataset which includes the contents of the speech files as text files[11]. It is an imbalanced dataset according to the number of files in each class. So, we used both the Unweighted Accuracy (UA) and the Weighted Accuracy (WA) as evaluation metrics. Also, to compare the output text of the ASR system and the Ground-Truth (GT) transcriptions of the reference data, the Word Error Rate (WER) and the Character Error rate (CER) metrics are used. Although it has been about 3 years since the ShEMO dataset became available, several research results have been reported on this dataset so far. During the research we did on this data, it became clear that some of the labels of these data are incorrect; therefore, in this paper, we will explain how to modify the ShEMO dataset.

The rest of the structure of this paper is as follows: In section 2, we review multimodal systems that have used text and audio information for SER and details about the ShEMO dataset. In section 3, we correct the errors in the ShEMO dataset with the help of the ASR system adapted for the Persian language. In section 4, we introduce the ASR-based SER system that works with the acoustic and linguistic information of the speech. In section 5, we examine the results and have a conclusion.

## II. RELATED WORKS

The traditional SER approach primarily includes two stages, which are known as feature extraction and feature classification[1], [6]. In the field of speech processing, researchers have obtained several features, including excitation source features, prosodic features, vocal tract contraction factors, and different combinations of features. The second stage includes the classification of features using classification methods such as machine learning (ML) and deep learning (DL) algorithms[6]. DL is considered an emerging research field in ML and has received more attention in recent years. DL techniques for SER have several advantages over traditional methods[6], [12], [13].

### A. Multimodal SER Using Audio and Text

Using acoustic features alone may not be enough for SER, as the speech presents messages that are beyond words. Words alone are not enough to convey a message; audio information is also required. Therefore, conveying meaning is not only related to how it is said (audio) but also to what is said (verbal)[2], [8], [14]. Eben et al. in [15] used low-level speech features and linguistic features for SER. They have performed the classification separately using speech and linguistic features, as well as combined using feature level fusion, and using BLSTM neural network. They showed that audio features perform better than linguistic features. However, the best results were obtained when both were combined. In [16], the concept of emotional salience was used to obtain linguistic information, linear classification and K-Nearest Neighbor (KNN) were used to obtain audio information, and finally, audio and linguistic information were combined. To combine audio and text features, at the decision level, it was assumed that these features were independent and logical OR function was used.

In [17], an algorithm based on belief networks was introduced to find emotional expression. It has been used to extract linguistic information from the output of an ASR based on the Hidden Markov Model (HMM) and a zero-gram LM, along with the confidence scores of each word. The goal was to find a probabilistic hypothesis that maximizes the posterior probability of a sequence of words, according to audio observations. Multi-Layer Perceptron (MLP) neural network with a 14-dimensional input feature vector and 7 output neurons was used to combine information. In [14], the transcriptions of an ASR system were used instead of the actual transcription of the sentences. Providing true transcription is costly and time-consuming, and the input to the system should only be speech signals. The results showed that the use of lower-quality transcription leads to less accuracy in separating classes that had the same level of arousal, but in the end, the audio and linguistic features are complementary to each other.

Research on two groups of children and the elderly is difficult due to the lack and difficulty of collecting data. In [10], a new dataset for the elderly was presented, which includes audio signals and speech transcriptions. New types of features such as BoAW[1] or word embedding vectors were introduced from sequence-to-sequence deep recurrent network architectures. In this work, the representation of audio features was based on the Fisher Vector (FV) encoding method. The linguistic information was used for the valance and the voice for the arousal. Then, a set of audio and linguistic features were extracted and tested, and finally, fusion strategies were investigated at the feature level and the decision level.

In [5], Peppino et al. have tried different methods to combine linguistic and acoustic information. Also, to obtain the word embedding vectors, BERT and GloVe methods were investigated and compared. Their results showed that the BERT embedding was a more suitable choice for representing linguistic information. They achieved 65.1% UA on the IEMOCAP dataset. In [8], 43 low-level audio features and 256-dimensional word embedding vectors were used for use in a bi-modal network. In [18], Wu et al. have used two separate methods including a Time-Synchronous Branch (TSB) and a Time-Asynchronous Branch (TAB) for emotion recognition. To get the correlation between each word and its corresponding audio, TSB combines the speech and text states in each frame of the input window and then merges them to form an embedding vector. On the other hand, TAB represents information between the sentences by combining the embeddings of some consecutive sentences in the text. The final emotion classification used both TSB and TAB embeddings. They were able to achieve a WA of 77.76% and a UA of 78.30% in recognizing the four emotions of happiness, sadness, anger, and neutral in the IEMOCAP dataset. In [9], audio features extracted from speech files were combined in an early fusion method with embedding vectors of words in the text corresponding to each sentence of the audio file, and 75.49% of WA was obtained on the IEMOCAP dataset. It should be mentioned that a pre-trained GloVe model was used to obtain the embedding vectors of the words in the text. In [19], using an ASR system based on the Wav2Vec2 model, information on the hidden layers of this model was extracted and injected into the audio information along with the text output of the speech file. For audio features, the MFCC method and the BERT model as text features are used. In this system, the text output of the sentences was obtained using a decoder based on the Connectionist Temporal Classification (CTC) algorithm. Also, the BLSTM network was used along with the attention mechanism for each of the audio and LMs, as well as for using the information of the hidden layers of the Wav2Vec2 model, and 63.4% of WA on the IEMOCAP dataset was obtained.

### B. Emotional Speech Databases

Many researchers use emotional speech databases in various research fields. The quality of the databases used and the performance obtained are the most important factors for evaluating an SER system. The available methods for collecting speech databases are different depending on the motivation of speech systems development[1], [3]. To develop emotional speech systems, speech databases are divided into three main types: (a) Simulated database, (b) Induced/Elicited database, and (c) Natural/Spontaneous database[1], [3], [4], [11]. In Table 1, a number of datasets of emotional speech available in the Persian language have been reviewed.

### C. Sharif Emotional Speech Database

The Sharif Emotional Speech Database (ShEMO), which was collected and published at Sharif University in 2018, includes 3000 semi-natural speech files that are equivalent to 3 hours and 25 minutes of speech samples collected from online radio broadcasts[11]. These files are in .wav format, 16-bit, 44.1kHz, and single-channel.

In this dataset, 87 people (including 31 women and 56 men), whose mother tongue is Farsi, were used to express the 5 main emotions of anger, fear, happiness, sadness, and surprise, as well as the neutral state without emotion. 12 annotators, including 6 men and 6 women, labeled these speech files and the voting method was used to determine the

TABLE I. PERSIAN EMOTIONAL SPEECH DATABASES

| Dataset | Publication year | Emotions | Number of Speech Files |
|---|---|---|---|
| Persian ESD[a] [20] | 2012 | fear, disgust, anger, happiness and sadness | 472 |
| SES[b] [21] | 2008 | neutral, surprise, happiness, sadness and anger | 1200 |
| PDREC[c] [22] | 2014 | anger, boredom, disgust, fear, neutral, surprise, and joy | 748 |
| ShEMO[d] [11] | 2018 | anger, fear, happiness, sadness, surprise and neutral | 3000 |

[a.] Persian Emotional Speech Database
[b.] Sahand Emotional Speech Database
[c.] Persian Drama Radio Emotional Corpus
[d.] Sharif Emotional Speech Database

---
[1] Bag-of-Audio-Words

final label. The native language of the annotators was Persian, and these people did not have any hearing or psychological problems. Their average age was 24.25 years and their standard deviation was 5.25 years in the age range of 17 to 33 years. Also, this dataset has been provided orthographically and phonetically according to the IPA[1] standard to be used to extract linguistic features[11]. The average duration of sentences is 4.11 seconds with a standard deviation of 3.41. Also, the text of each sentence is placed in a file in. ort format. Information about the distribution of classes as well as the statistical information related to the files of the ShEMO dataset is shown in Table 2.

Using the GT text transcriptions of a dataset to combine linguistic and acoustic features can lead to the high accuracy of the SER model in the desired dataset. Table 3 contains information about the text of the sentences related to each speech file of the ShEMO dataset. This information includes the number of tokens or the number of words in each emotional class, the number of unique words in each class, and the number of words that exist exclusively in a certain emotional class but have not appeared in other classes.

In the following, we review the work done on the ShEMO dataset. In [23], different audio features along with several classification methods have been tested on 17 datasets. In this paper, the effect of using 9 classifiers and 17 sets of audio features on datasets has been investigated in a speaker-independent speech emotion recognition system. Various classic algorithms such as random forest, SVM, and neural networks were tested. Also, a set of various low-level and high-level features extracted using the openSMILE tool, along with BoAW-based audio features and neural network-based features were examined in various experiments. Finally, a UA of 64% is reported for the ShEMO dataset using a system based on the wav2vec model. These tests have been done in the form of speaker-independent cross-validation. In [24], the effect of using different loss functions in capsule convolutional networks on spectrograms has been investigated. It should be mentioned that the main application of these networks is in image processing and detection of rotation or transfer in images, which normal convolutional networks are not able to detect such cases. Also, data augmentation techniques such as additive noise and VTLP[2] were used and finally, they achieved a WA of 71.43% on the ShEMO dataset. In [25], A 1D convolutional neural network (CNN) was used on MFCC features for SER in the ShEMO dataset and achieved 74% WA.

In [13], different DL models were tested on various Low-Level Descriptors (LLD) and functional acoustic features. UA of 65.20% was obtained using a CNN network on the emo_large feature set. In [7], using the word embedding vectors obtained from the Persian *fastText* model, 73.73% UA was obtained by fusing CNN models on GT text transcription embeddings and acoustic features using the early fusion method.

Other works related to SER done on the ShEMO dataset include multilingual anger identification from MFCC features with CRNN in [26], a baseline for unsupervised cross-lingual SER in [27], a multilingual benchmark for SER with SERAB used to evaluate a range of recent baselines in [28], a semi-supervised learning approach for cross-lingual SER in [29], the influence of speech features on the recognition of anger and neutral emotions in different languages with the pitch, intensity, formants and MFCCs features in [30], and investigation of human behavior regarding the perception of emotions in speech with SVM in a cross-cultural study in [31].

To the best of our knowledge, the use of ASR output sentences for the ShEMO dataset has not been reported so far. In Shahid Beheshti University's Intelligent Sound Processing Laboratory (ISP-Lab), for the first time, we used the text of the output sentences of a Persian ASR system to combine audio and text information to recognize the emotions of the ShEMO dataset. By comparing the output sentences of ASR and GT sentences, we noticed some contradictions and errors in this dataset. These errors are related to the contradictions of the sentences in the speech (.wav) and text (.ort) files and also the contradictions of the labels. In the continuation of our project, the correction of the errors in the ShEMO dataset and also the implementation of a Persian ASR-based SER system has been addressed. We also put the corrected data on GitHub[3] for public research.

### III. SHEMO MODIFICATION

#### A. *Persian ASR based on the wav2vec2 model*

As the speech recognition system, the wav2vec2-large-xlsr-persian model is used, which is trained on the CommonVoice's Farsi samples[32]. The decoder of this system is a CTCBeamDecoder. It can predict the best sentence by using the beam search among the scores related to the characters in each speech frame[33].

TABLE II. STATISTICS OF SHEMO UTTERANCES

| Emotional State | Number | | | Duration | | | |
|---|---|---|---|---|---|---|---|
| | *Female* | *Male* | *Total* | *Min* | *Max* | *Mean* | *SD*[a] |
| Anger | 455 | 604 | 1059 | 0.44 | 22.42 | 3.61 | 2.63 |
| Fear | 22 | 16 | 38 | 0.76 | 8.97 | 3.17 | 1.84 |
| Happiness | 111 | 90 | 201 | 0.82 | 13.39 | 3.81 | 2.36 |
| Neutral | 284 | 744 | 1028 | 0.56 | 33.32 | 4.89 | 4.1 |
| Sadness | 271 | 178 | 449 | 0.69 | 27.89 | 4.84 | 3.7 |
| Surprise | 120 | 105 | 225 | 0.35 | 10.95 | 1.79 | 1.45 |
| Total | 1263 | 1737 | 3000 | 0.35 | 33.32 | 4.11 | 3.41 |

a. Standard Deviation

TABLE III. SHEMO TEXT INFORMATION

| Emotions | # Tokens | # Unique Words | # Class-Specific Words |
|---|---|---|---|
| Anger | 11956 | 3275 | 1698 |
| Fear | 255 | 154 | 14 |
| Happiness | 2026 | 969 | 303 |
| Neutral | 14723 | 4089 | 2434 |
| Sadness | 4351 | 1501 | 510 |
| Surprise | 968 | 432 | 91 |
| Anger | 11956 | 3275 | 1698 |

---

[1] International Phonetic Alphabet
[2] Vocal Tract Length Perturbation
[3] https://github.com/aliyzd95/ShEMO-Modification

Having contextualized audio classifications and no alignment problems, Wav2Vec2 does not require an external LM or dictionary to yield acceptable audio transcriptions. However, the results clearly show that using Wav2Vec2 in combination with a suitable LM can yield a significant improvement.

### B. Persian Colloquial Corpora

Since the content of the sentences of the ShEMO dataset is conversational, we build n-gram LMs using the KenLM tool on different Persian colloquial text corpora (see Table 4):

- **LSCP:** This corpus includes 120 million Persian conversational sentences from 27 million tweets, which is accompanied by a derivation tree, grammatical tagging (POS), sentiment polarity, and translation of each sentence in five languages: English, German, Czech, Italian and Hindi[34].

- **MirasOpinion:** According to the collectors, this dataset was the largest sentiment analysis dataset in Farsi until its release. The number of comments from 2.5 million comments on the DigiKala website has been reduced to one million comments after a series of pre-processing and then human resources have been used to label them. The total number of documents is more than 93 thousand, which are categorized into 3 classes: positive, negative, and neutral[35].

- **DK dataset-2:** Real transaction data of more than 2 million customers and 100,000 products have been sampled. This data contains one hundred thousand examples of user comments, which include several comments for the same product.

- **W2C:** This corpus includes text corpora in 120 different languages, which are automatically collected from web pages and Wikipedia[36].

From Table 4, it can be seen that the W2C corpus, in addition to the fact that the number of words not covered in the ShEMO text corpus is only 235 tokens, with a reasonable size, covers a high percentage of different words regardless of the ShEMO corpus. Also, the WER metric for the sentences predicted by the speech recognition system, using the LM on this corpus, is lower than others.

### C. Contradictions Correction

As mentioned, the ShEMO dataset contains 3000 audio files along with 3000 text files for each sentence as their GT transcription. The text file of the sentence related to the corresponding audio file can be found through the names of the files. In fact, the audio and the text file of an utterance have the same name. But out of 3000 files, only 2838 have the same name. Upon further investigations, we found that some of these text files have the wrong names and referred to the wrong audio file.

In Fig. 1, examples of errors in referencing audio and text files can be seen. To recognize the correct names of these files, a 5-gram LM is used on the corpus of the ShEMO dataset. Thus, the Persian ASR with the help of the LM based on the ShEMO tokens will recognize sentences more accurately. Then the sentences whose WER and CER metrics are more than a threshold, here is 0.5, are compared with each other to find the correct text file. There are 347 files out of a total of 3000 files in the dataset that meet the mentioned conditions. The result of modifying the dataset is much less errors in sentence recognition. The pseudo-code related to finding the correct sentences for each audio file is shown in Fig. 2. As shown in Table 5, the WER in this dataset has been greatly reduced after correcting the text files related to each audio file.

### D. Labels Correction

Various tests have been performed to prove the existence of errors in the ShEMO dataset, as well as tests to prove the correctness of the dataset after modification. After modifying the dataset, it was found that there are 163 files whose audio file labels are different from their text file labels. Table 6 shows the result of testing a CNN on the new dataset. In this experiment, these 163 files have been used as the test set and the rest of the files have been used to train the model. The better performance of the model by choosing the text file label as the final label of these 163 files shows that the text file label is the correct label for these files.

TABLE IV. PERSIAN COLLOQUIAL CORPORA INFORMATIONS

| Corpus | Source | # Tokens | # OOV words | % ShEMO Coverage | Size | WER% |
|---|---|---|---|---|---|---|
| LSCP | Twitter | 307 million | 175 | 66% | 2.5 GB | 62.03 |
| MirasOpinion | Digikala Website | 3.4 million | 1588 | 71% | 29 MB | 62.40 |
| DK Dataset-2 | Digikala Website | 2.8 million | 1685 | 74% | 24 MB | 62.67 |
| W2C | Web Pages | 125 million | 235 | 56% | 980 Mb | **58.60** |

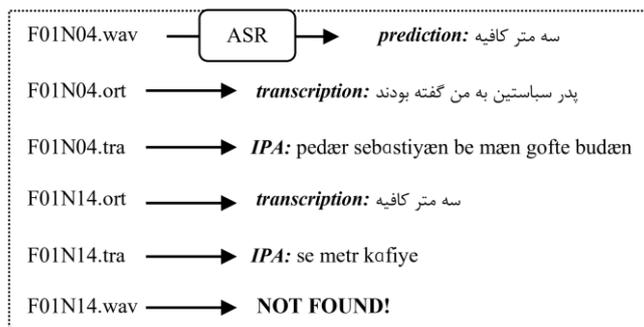

Fig. 1. Example of the ShEMO Errors

```
1 FIND CANDIDATE SENTENCES AND PREDICTIONS (Condition: WER>0.5 AND CER>0.5)
2 FindBestMatching (sentenceList, predictionList)
    for each sentence in sentenceList do
            werList <- CalculateWER (sentence, PredictionList)
            bestMatchingList <- FindMinimum(werList)
            if bestMatchingList.length is 1:
                    Correction(sentence.id, bestMatchingList.element.id)
                    unusedSentenceList.delete(sentence)
                    unusedPredictionList.delete(bestMatchingList.element)
            else
                    unusedSentenceList.add(sentence)
                    unusedPredictionList.add(werList.elements)
3 FindBestMatching (unusedSentenceList, unusedPredictionList)
4 if unusedSentenceList.isEmpty() and unusedPredictionList.isEmpty()
            Exit
5 goto 3
```

Fig. 2. ShEMO Modification Pseudo-code

TABLE V. WER BEFORE AND AFTER SHEMO MODIFICATION

| Language Model | ShEMO Modification | WER% |
|---|---|---|
| 5-gram LM on ShEMO Corpus | Before | 34.62 |
|  | After | 14.71 |
| 4-gram LM on W2C | Before | 51.97 |
|  | After | 30.79 |

TABLE VI. LABELS SELECTION EXPERIMENTS USING A CNN

| Acoustic Model | Select Labels | WA% | UA% |
|---|---|---|---|
| CNN (1D) | .wav files | 33.55 | 22.55 |
|  | .ort files | 72.25 | 43.92 |

In the following, tests based on the reference paper of the ShEMO dataset have been conducted to prove the correctness of the new labels. At first, the Support Vector Machine (SVM) model presented in the paper has been checked and tested on the modified dataset. It should be noted that the implementation of this model is completely consistent with what was explained in the ShEMO paper. The result of testing the SVM model on the old dataset, before modification, is almost equal to the result presented in the ShEMO paper. But after modifying the labels of the dataset, as it is clear in table 7, the UA of the model has increased by about 5%, which indicates that a better result has been obtained by modifying the labels of the dataset. In addition, both in the section on correcting the references of the text files to the audio files and in the section on correcting the labels related to the samples, all the cases have been checked manually and the correctness of the modification of the ShEMO dataset can be confirmed.

## IV. ASR-BASED SER

Using GT transcriptions to combine linguistic and acoustic features, can lead to high model accuracy in emotion recognition. But when such a product leaves the lab environment, a transcript of the spoken utterances of the individuals will not be available. For this reason, a speech-to-text or ASR system should be used to obtain the text transcriptions of the sentences[14], [19]. Fig. 3 shows the proposed ASR-based Persian SER system.

### A. Acoustic Model of SER

To extract high-level acoustic features, the openSMILE toolkit is used, which provides a set of different features for extracting from speech files[37]. The emo_large feature set, which extracts the largest number of high-level features from each audio file, includes 6552 features obtained by applying high-level statistical functions to LLDs. These features have been obtained by applying 39 statistical functions to 56 LLDs along with their 56 deltas. A CNN takes the feature vector of each sentence as input and passes through the 1D convolution layers with the rectifier activation function (*ReLU*) and max pooling layers. Also, batch normalization and dropout were used to prevent overfitting. Finally, after a global mean pooling layer, the *softmax* function determines the probability of each emotional class.

### B. Linguistic Model of SER

The *fastText* model is a widely used word embedding model pre-trained on large text corpora including Wikipedia and Common Crawl[38]. In this paper, a *fastText* model trained on the OSCAR dataset, which is a multilingual corpus, is used. This model embeds a 100-dimensional vector for each word and contains more than 4 million unique words[39].

TABLE VII. BASELINE SVM MODEL WITH CORRECTED LABELS

| Machine Learning Model | ShEMO Modification | WA% | UA% |
|---|---|---|---|
| SVM | After | **76.65** | **63.62** |
|  | Before | 72.95 | 58.66 |
|  | ShEMO Paper [11] | NA | 58.02 |

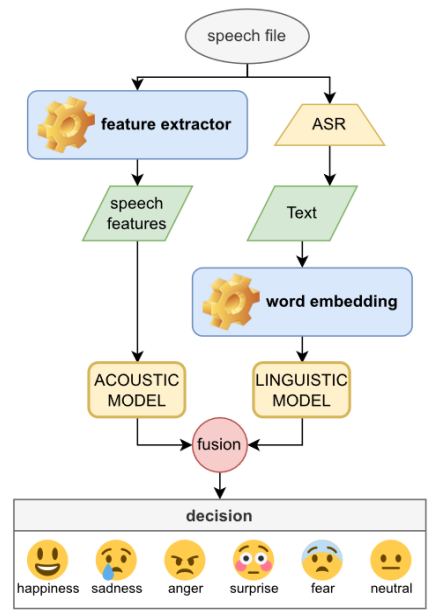

Fig. 3. Proposed ASR-based Persian SER system

Here, each sentence is tokenized first and the length of the sentences is obtained. The largest sentence in this dataset contains 68 tokens and the rest of the sentences are zero-padded to be the same length as the largest sentence. Finally, the 100-dimensional vectors related to each word in the sentence are extracted from the *fastText* model and the weights of the embedding layer in the desired neural network are determined. The neural network used to use word embedding vectors to identify emotion in the sentence includes parallel 2D convolution layers with different numbers of filters. Finally, these features concatenate with each other and the *softmax* function calculates the probability of occurrence of each emotional class.

*C. Fusion Model*

With an early fusion method, the outputs of the acoustic and linguistic-based SER models, before reaching the layer corresponding to the *softmax* function and calculating the probability of the output class, are concatenated together and a feature vector that includes audio and linguistic features is obtained[5]. Then this vector is given as an input to a deep neural network (DNN) consisting of fully connected layers and finally, after passing through the dense layers, the probability output of each class will be calculated. Fig. 4 shows how the combination system works.

V. RESULTS AND CONCLUSION

We implemented a 5-fold cross-validation for our experiments. It should be noted that due to the small number of files with fear labels (38 files in total), these files are removed from the experiments of this research.

In this section, we examine the result of fusing audio and text information for SER using the early fusion method. As expected, the best results are obtained when using GT transcripts as linguistic information. However, as mentioned in the previous sections, the exact text transcriptions of the sentences expressed as reference data are not available outside the laboratory environment, and to obtain the text transcripts of the utterances, an ASR system must be used for real-time applications. In fact, we are only allowed to use information from the speech files for SER.

Using the output sentences of the speech recognition system, they have also helped to improve the performance of the final SER system. In fact, in this model, only audio information is used to identify the emotion, because the ASR system also uses audio information to obtain the text of the utterances. As shown in Table 8, the fused ASR-based model provides better performance (69.73%) than the acoustic model (66.12%) and has improved UA by 3.61%.

In conclusion, we used a Persian speech-to-text system to obtain textual information from speech files. Also, in this paper, different Persian colloquial text corpora were investigated to improve the performance of the Persian ASR system. This ASR system also revealed some errors and inconsistencies in the ShEMO dataset, which were also corrected in this paper.

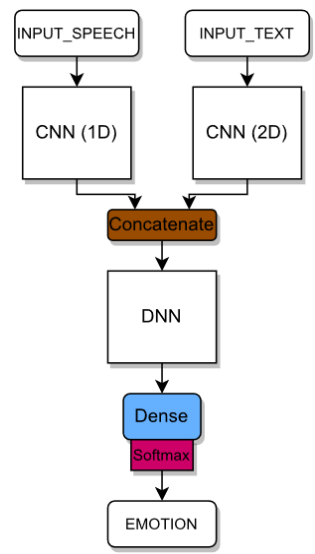

Fig. 4. Audio and Text Early-Fusion

TABLE VIII. COMPARISON OF THE RESULTS

| Model | Used Features | WA% | UA% |
|---|---|---|---|
| SVM (Baseline [11]) | eGeMAPS | 76.65 | 63.62 |
| CNN (1D) (Acoustic Model) | emo_large | 79.68 | 66.12 |
| CNN (2D) (Linguistic Model) | fastText (GT) | 58.01 | 51.37 |
| Early-Fusion Model (DNN) | emo_large + fastText (GT) | **81.60** | **74.68** |
| | emo_large + fastText (ASR transcriptions) | **80.51** | **69.73** |

Finally, by modifying the ShEMO dataset, the WER metric of the speech recognition system with the help of the LM on the W2C corpus was reduced from 51.97 to 30.79. Also, the fusion of text and audio information for emotion recognition was investigated and the best performance was obtained when using GT text transcriptions of the reference data. It can be concluded that by using an adapted speech recognition system, sentences with less errors can be obtained, which will improve the performance of the SER when only audio information is used.